\documentclass[twocolumn,showpacs,amssymb,aps,prc,nofootinbib,floatfix,superscriptaddress]{revtex4-1}

\usepackage{supertabular}
\usepackage{lipsum}
\usepackage{blindtext}
\usepackage{amsfonts}
\usepackage{tensor}
\usepackage{graphicx}
\usepackage{pict2e}
\usepackage{balance}
\usepackage{blindtext}

\usepackage{amssymb, amsmath,environ}

\usepackage{color}
\usepackage[colorlinks=true,urlcolor=black,linkcolor=black,citecolor=red]{hyperref}


\newcommand{\bse}{\begin{subequations}}
	\newcommand{\ese}{\end{subequations}}
\newcommand{\be}{\begin{equation}}
\newcommand{\ee}{\end{equation}}

\NewEnviron{eqsplit}{%
	\begin{equation}\begin{split}
	\BODY
	\end{split}\end{equation}
}

\makeatletter
\newcommand*\bigcdot{\mathpalette\bigcdot@{.5}}
\newcommand*\bigcdot@[2]{\mathbin{\vcenter{\hbox{\scalebox{#2}{$\m@th#1\bullet$}}}}}
\makeatother

\newcommand{\bea}{\begin{eqnarray}}
\newcommand{\eea}{\end{eqnarray}}
\newcommand{\ba}{\begin{array}}
	\newcommand{\ea}{\end{array}}

\newcommand{\nn}{{\nonumber}}

\newcommand{\la}{\langle}
\newcommand{\ra}{\rangle}

\def \la{\langle}
\def \ra{\rangle}
\def \sNN{\sqrt{s_{NN}}}
\def \beq{\begin{equation}}
\def \eeq{\end{equation}}
\def \beqa{\begin{eqnarray}}
\def \eeqa{\end{eqnarray}}
\def \vd{\vec{v}_{\text{drift}}}

\begin{document}

\title{Charge dependent directed flow splitting from baryon inhomogeneity and electromagnetic field}

\author{Tribhuban Parida}
\email[]{tribhu.451@gmail.com}
\affiliation{Department of Physical Sciences, Indian Institute of Science Education and Research Berhampur, Transit Campus (Govt ITI), Berhampur-760010, Odisha, India}

\author{Sandeep Chatterjee}
\email[]{sandeep@iiserbpr.ac.in}
\affiliation{Department of Physical Sciences, Indian Institute of Science Education and Research Berhampur, Transit Campus (Govt ITI), Berhampur-760010, Odisha, India}

\author{Subhash Singha}
\email[]{subhash@impcas.ac.cn}
\affiliation{Institute of Modern Physics Chinese Academy of Sciences, Lanhzou 730000, China}

\begin{abstract}
This work aims to understand the recent experimental data from the STAR collaboration on the 
system size dependence of directed flow splitting between oppositely charged hadrons \cite{Taseer:2024sho}. 
Previously, we have studied the role of baryon inhomogeneity on charge dependent directed 
flow. We now incorporate the effects of the electromagnetic (EM) field albeit perturbatively, 
as implemented in Ref. \cite{Gursoy:2018yai}. This enables us to compare the relative 
contributions between baryon inhomogeneity and EM field on charge dependent directed flow. 
Our model calculation describes the experimental data on the centrality and system size dependence of the mid-rapidity directed flow slope splitting, $\Delta dv_1/dy$, between protons and anti-protons. Our results indicate that in central collisions, where the EM field strength is negligible, the inclusion of EM field effects does not influence the splitting between protons and anti-protons. This suggests that the observed system size dependence of $\Delta dv_1/dy (p-\bar{p})$ in central collisions arises solely from enhanced baryon stopping in larger collision systems. However, in semi-central and peripheral collisions, both baryon diffusion and EM field effects contribute to the splitting. Furthermore, the centrality dependence of $\Delta dv_1/dy (p-\bar{p})$ is highly sensitive to the electrical conductivity of the medium, making it a potential probe for extracting this transport coefficient in the QCD medium through model-to-data comparisons. However, achieving this requires a precise determination of the background baseline originating from baryon diffusion. Additionally, further investigation is needed to understand $\Delta dv_1/dy$ for oppositely charged kaons and pions, particularly by incorporating the diffusion of other conserved charges. 
\end{abstract}


\maketitle

\section{Introduction}

Flow coefficients characterize the azimuthal anisotropy of hadrons produced in relativistic heavy-ion collisions \cite{Ollitrault:1992bk,Voloshin:2008dg}. Among them, the directed flow ($v_1$) is defined as the coefficient of the first-order harmonic in the Fourier expansion of the azimuthal distribution of final-state hadrons relative to the reaction plane angle ($\Psi_{RP}$):
\beq
\frac{dN}{p_T dp_T dy d\phi} = \frac{dN}{p_T dp_T dy} \left[ 1 + 2 v_1(p_T,y) \cos(\phi-\Psi_{RP}) + ... \right].
\eeq
Due to the geometry of the collision, $v_1$ is an odd function of rapidity ($y$), and its magnitude is typically expressed in terms of the mid-rapidity slope, $dv_1/dy$ \cite{Snellings:1999bt,STAR:2011hyh,STAR:2014clz,STAR:2017okv}. At LHC and the highest RHIC energies, measurements of the directed flow for charged particles exhibit a negative mid-rapidity slope \cite{STAR:2005btp,STAR:2008jgm,PHOBOS:2005ylx,STAR:2011gzz,ALICE:2013xri}. Hydrodynamic model calculations incorporating a tilted initial energy density profile successfully capture this feature \cite{Bozek:2010bi}.

In recent years, extensive experimental studies have been conducted on the directed flow of identified hadrons across a broad range of collision energies \cite{STAR:2017okv,STAR:2011hyh,STAR:2014clz,STAR:2016cio,STAR:2025rtd,STAR:2023jdd}. These measurements are often presented by comparing the directed flow of particles with similar masses but different conserved charge quantum numbers, such as $\pi^+ - \pi^-$, $K^+ - K^-$, and $p - \bar{p}$ \cite{STAR:2025rtd,STAR:2023jdd,STAR:2017okv,STAR:2014clz}. While $\pi^+$ and $\pi^-$ share the same mass but differ in electric charge, $K^+$ and $K^-$ also have identical masses but differ in both electric charge and strangeness. Similarly, protons and anti-protons have the same mass but differ in both baryon number and electric charge.

So far, two primary mechanisms have been identified as potential sources of directed flow splitting for particles of equal mass:
\begin{enumerate}
    \item The presence of nonzero conserved charge density in the medium \cite{Bozek:2022svy,Parida:2022ppj,Parida:2022zse,Jiang:2023fad}.
    \item The influence of the electromagnetic field \cite{Gursoy:2014aka, Nakamura:2022ssn, Benoit:2025amn,Inghirami:2019mkc,Dubla:2020bdz,Sun:2020hvb,Sun:2023adv,Chatterjee:2018lsx,Das:2016cwd}.
\end{enumerate}

In hydrodynamic models, it has been demonstrated that an inhomogeneous distribution of net baryon density in the fireball leads to a splitting of directed flow between baryons and anti-baryons \cite{Bozek:2022svy,Parida:2022ppj,Parida:2022zse,Jiang:2023fad}. Notably, this $v_1$ splitting is highly sensitive to the initial distribution of net baryons in the medium \cite{Bozek:2022svy,Parida:2022ppj,Parida:2022zse,Jiang:2023fad,Du:2022yok}. Model calculations further indicate that baryon diffusion plays a significant role in influencing this splitting \cite{Parida:2023rux,Parida:2023ldu}. Consequently, it has been proposed that a model-to-data comparison of baryon–anti-baryon $v_1$ splitting could provide insights into the baryon stopping mechanism and serve as a means to constrain the baryon diffusion coefficient—an essential transport property of the QCD medium \cite{Bozek:2022svy, Parida:2022zse, Parida:2022ppj, Du:2022yok, Parida:2023rux}. Just as net-baryon inhomogeneity leads to splitting in baryon–anti-baryon observables, an inhomogeneous distribution of net strangeness and net electric charge in the medium is also expected to induce a similar splitting between hadron pairs of equal mass, such as $K^+ - K^-$ and $\pi^+ - \pi^-$.

On the other hand, the electromagnetic field generated by positively charged spectators also contributes to the splitting of $v_1$ ($\Delta v_1$) between oppositely charged hadrons \cite{Dubla:2020bdz,ALICE:2019sgg,STAR:2023jdd,STAR:2023wjl,STAR:2025rtd,STAR:2016cio,STAR:2017ykf,ALICE:2012nhw,Gursoy:2018yai,Gursoy:2014aka,Panda:2025lmd,Sun:2020hvb,Sun:2023adv,Chatterjee:2018lsx,Das:2016cwd,Benoit:2025amn,Nakamura:2022ssn,Nakamura:2022idq}. The time dependent magnetic field in the expanding QGP medium gives rise to a net electric current along the impact parameter direction due to Faraday induction, the Lorentz force, and the Coulomb force \cite{Gursoy:2014aka, Gursoy:2018yai}. This induced electric field exerts a sideward force, pushing positively and negatively charged constituents of the QGP medium in opposite directions. 
As a result, this force manifests in
the momentum-space distribution of the finally produced
charged hadrons and can be quantified by measuring
the $\Delta v_1$ between oppositely charged
hadrons. Previous studies have shown that the $\Delta v_1$ between oppositely charged hadrons is highly sensitive to the electrical conductivity ($\sigma$) of the medium \cite{Benoit:2025amn, Gursoy:2018yai}. Similar investigations have also explored the impact of electrical conductivity on oppositely charged heavy-flavored hadrons \cite{Chatterjee:2018lsx}. Therefore, to constrain the electrical conductivity of the strongly interacting QCD medium, it is essential to both measure and theoretically study charge-dependent $v_1$ splitting.

Measurements by the STAR collaboration have examined the splitting of the mid-rapidity directed flow slope, $\Delta dv_1/dy$, for pairs such as $\pi^+-\pi^-$, $K^+-K^-$, and $p-\bar{p}$ across different centralities \cite{STAR:2023jdd}. These measurements reveal a sign change in $\Delta dv_1/dy$ from central collisions, where the EM field is weaker, to peripheral collisions, where the EM field is stronger. This observed sign reversal has been attributed to the influence of the initial EM field. However, in our previous study, we argued that the centrality dependence of directed flow splitting may not be solely driven by the EM field. The presence of nonzero conserved charge and its subsequent diffusion in the medium can play a crucial role in shaping the centrality dependence of $\Delta dv_1/dy$ \cite{Parida:2023ldu,Parida:2023rux}. In particular, we demonstrated that the sign change in $\Delta dv_1/dy (p-\bar{p})$ from central to peripheral collisions can be reproduced by incorporating nonzero baryon diffusion in our model, even without considering any EM field effects \cite{Parida:2023ldu}. This suggests that both conserved charges diffusion and the EM field contribute to the observed $v_1$ splitting for these hadron pairs. To isolate the signal of the EM field, it is essential to disentangle it from the background contributions arising from conserved charge dynamics.

\begin{figure}[th!]
	\includegraphics[scale=0.5]{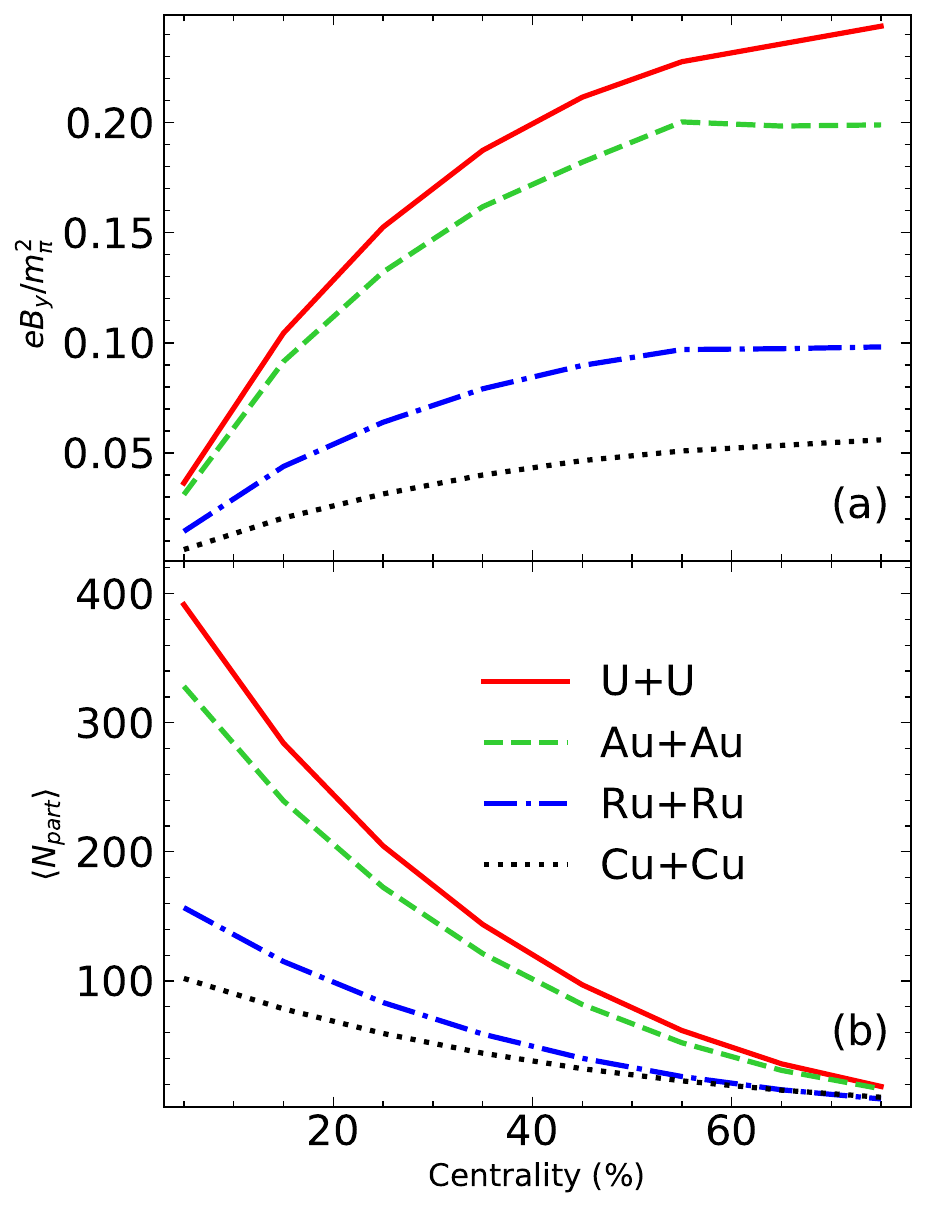} 
	\caption{(a) Centrality dependence of the $y$-component of the initial magnetic field ($eB_y$) in U+U, Au+Au, Ru+Ru, and Cu+Cu collisions at $\sNN=200$ GeV. The magnetic field is calculated at $ (x, y, z) = (0,0,0) $ at a proper time of $ \tau = 0.6 $ fm by solving Maxwell’s equations, assuming an electrical conductivity of $\sigma = 0.023 $ fm$^{-1}$. The values are presented in units of $ m_{\pi}^2 $. The strength of $ eB_y $ increases with system size for a given centrality. A pronounced system-size dependence is observed in peripheral collisions, whereas in central collisions, the field strength remains very small across all systems.  
(b) The average number of participants as a function of centrality for U+U, Au+Au, Ru+Ru, and Cu+Cu collisions at $ \sNN = 200 $ GeV, obtained using the Monte Carlo Glauber model. For a given centrality, baryon stopping increases with system size. Notably, a clear system-size dependence of baryon stopping is observed in central collisions.}
	\label{fig:eBy_AuRuCu}
        \end{figure}

Recently, the STAR collaboration presented centrality-dependent measurements of $\Delta dv_1/dy$ between oppositely charged hadrons for $\sNN = 200$ GeV across different collision systems, including U+U, Au+Au, and Zr+Zr \cite{Taseer:2024sho}. A comparison of these measurements revealed a pronounced system size dependence of $\Delta dv_1/dy$, with the effect being most significant for $p-\bar{p}$. At a given centrality, the number of spectator nucleons varies across different collision systems, directly affecting the strength of the generated electromagnetic (EM) field. To illustrate this, Fig. \ref{fig:eBy_AuRuCu} (a) presents the model calculation of the initial $eB_y$ produced by spectator nucleons at the position $(x, y, \eta_s) = (0, 0, 0)$ at proper time $\tau=0.6$ fm for Cu+Cu, Ru+Ru, Au+Au and U+U collisions across different centralities. This demonstrates that the EM field strength increases with system size, leading to a stronger influence on the created QGP medium. Consequently, the observed system size dependence of $\Delta dv_1/dy$ at $\sNN=200$ GeV could serve as a potential signature of the EM field.

However, it is important to recognize that as the system size increases, the initial baryon deposition in the medium also increases. To illustrate this, Fig. \ref{fig:eBy_AuRuCu}(b) presents the average number of participants, $\la N_{part} \ra$, as a function of centrality for different collision systems. The results reveal a clear system size dependence in the initial baryon stopping at a given centrality. This indicates that the observed system size dependence of $\Delta dv_1/dy (p-\bar{p})$ may not be solely attributed to the EM field but could also stem from variations in the amount of initial baryon deposition. Consequently, disentangling the effects of baryon stopping from those of the EM field is crucial for accurately interpreting $\Delta dv_1/dy(p-\bar{p})$ as a signature of electromagnetic effects.

Furthermore, an intriguing feature can be observed in Fig. \ref{fig:eBy_AuRuCu}. In central collisions, the strength of the EM field is minimal, whereas baryon stopping is at its maximum, exhibiting a clear system size dependence in $\la N_{part} \ra$. In contrast, in peripheral collisions, where baryon stopping is significantly reduced and shows little system size dependence, the strength of $eB_y$ displays a pronounced system size dependence. This contrasting behavior highlights the need to explore the interplay between baryon stopping and EM field effects across different centrality regions. The primary objective of this work is to investigate this interplay using the charge-dependent splitting of directed flow.

\section{Framework}
In this phenomenological study, we employ a hydrodynamic model to simulate the evolution of the QGP medium. The publicly available MUSIC code is utilized \cite{Denicol:2018wdp,Schenke:2010nt}, incorporating the evolution of net baryon density alongside the energy-momentum tensor. The detailed implementation of the net baryon evolution equation and its diffusion within MUSIC is provided in Ref. \cite{Denicol:2018wdp}. The initial conditions for energy and net baryon density in the hydrodynamic evolution are adopted from Ref. \cite{Bozek:2010bi, Parida:2022ppj, Parida:2022zse}. Additionally, the effect of the electromagnetic (EM) field is implemented in a manner similar to the approach by Gurósoy et al. in Ref. \cite{Gursoy:2014aka, Gursoy:2018yai}. For a comprehensive information about the initial conditions, hydrodynamic evolution with baryon diffusion, and the incorporation of EM field effects within this framework, the reader is referred to the aforementioned references. However, for completeness, we provide a brief description in this section.

For a given centrality, we construct a smooth transverse profile of participant and binary collision sources by averaging over $10^4$ Monte Carlo (MC) Glauber events. In each MC Glauber event, all participant and binary collision sources are rotated by the second-order participant plane angle ($\Psi_2^{PP}$) of the transverse energy distribution, aligning $\Psi_2^{PP}$ along the positive x-direction \cite{Shen:2020jwv}. The sources are then smeared using a two-dimensional Gaussian distribution with a width of 0.4 fm.
With this setup, we define the coordinate system such that the averaged initial condition has $\Psi_2^{PP} = 0$, with the impact parameter vector pointing in the positive x-direction. Similarly, a smooth spectator charge density distribution is obtained by averaging over multiple MC Glauber events. These smooth spectator profiles are then used to compute the spatiotemporal evolution of the electromagnetic field \cite{Gursoy:2018yai}.

We construct a three-dimensional smooth initial profile of energy and net baryon density by using the transverse distributions of participant and binary collision sources with a parameterized rapidity envelope profile. The rapidity envelope extend them along the space-time rapidity direction. For the transverse energy deposition, we employ the two-component energy deposition model and adopt the tilted initial condition model to establish the initial three-dimensional energy density profile. The energy density at a constant proper time $\tau_0$, which serves as the input for hydrodynamic evolution, follows the same form as used in our previous study in Ref. \cite{Parida:2022ppj, Parida:2022zse}. In this tilted initial condition model, a free tilt parameter ($\eta_m$) controls the tilt of the energy distribution in the reaction plane \cite{Bozek:2010bi}. This parameter is appropriately chosen to reproduce the directed flow of charged hadrons. A single-shot hydrodynamic evolution is then performed using the event-averaged three-dimensional initial condition for a given centrality.

Furthermore, for the initial net-baryon distribution, we utilize our previously proposed two-component baryon deposition model, which has proven to be highly effective in reproducing the directed flow of identified hadrons across a broad range of collision energies. The initial net-baryon density is given by \cite{Parida:2022ppj,Parida:2022zse}:

\beqa
  n_{B} \left( x_{\perp}, \eta_s; \tau_0 \right) &=& N_{B} \left[ \left( N_{+}(x_{\perp}) f_{+}^{n_B}(\eta_{s}) + N_{-}(x_{\perp})f_{-}^{n_B}(\eta_{s})  \right)\right.\nn\\
                           &&\left.\times \left( 1- \omega \right) + N_{bin} (x_{\perp}) f^{n_B}_{\text{bin}}\left(\eta_{s}\right) \omega \right] 
 \label{eq:two_component_baryon_profile}    
\eeqa
Here, $N_{+}(x_{\perp})$ and $N_{-}(x_{\perp})$ represent the participant densities of the nuclei moving in the positive and negative rapidity directions, respectively. The term $N_{bin}(x_{\perp})$ accounts for the contribution from binary collision sources at each point in the transverse plane. The parameter $\omega$ is a free parameter that determines the relative contribution of participant and binary collision sources. This parameter, $\omega$, effectively controls the tilt of the baryon distribution in the reaction plane \cite{Parida:2022ppj}. Its value, along with $\eta_m$, is chosen appropriately to reproduce the rapidity dependence of the directed flow of pions, protons, and anti-protons \cite{Parida:2022ppj,Parida:2022zse}.

The functions $f_{\pm}^{n_B}(\eta_s)$ represent parameterized rapidity envelope profiles that describe the asymmetric baryon deposition by forward- and backward-moving participants. These profiles are taken from Ref. \cite{Denicol:2018wdp}. The free parameters in these rapidity envelopes are constrained by comparison with experimental data on the rapidity dependence of net-proton yields \cite{Du:2022yok,Parida:2022ppj}. Additionally, we introduce a forward-backward symmetric rapidity envelope profile, which is multiplied by the $N_{coll}$ sources. The symmetric profile is defined as $f_{bin}^{n_B}(\eta_s)  = f_{+}^{n_B}(\eta_{s}) + f_{-}^{n_B}(\eta_{s})$. The normalization factor $N_B$ in Eq.~\ref{eq:two_component_baryon_profile} is not treated as a free parameter; instead, it is determined by the total net baryon number initially carried by the participants, following the constraint: $ \int  \tau_{0}  n_{B} \left( x_{\perp}, \eta_s, \tau_{0} \right) dx_{\perp}  d\eta_s  = N_{\text{part}}$ as taken in Ref. \cite{Denicol:2018wdp}.

The hydrodynamic evolution of these initial profiles is carried out using the MUSIC code \cite{Denicol:2018wdp}, where the initial velocity field follows the Bjorken flow ansatz. During the evolution, we employ the NEoS-BQS equation of state \cite{Monnai:2019hkn}, which ensures both strangeness neutrality and a fixed baryon-to-charge density ratio within each fluid cell. In this study, we assume a constant specific shear viscosity ($c_\eta = \frac{\eta s}{\epsilon + p} = 0.08$) and neglect contributions from bulk viscosity. For baryon diffusion in the hydrodynamic evolution, we adopt the following form of the baryon diffusion coefficient ($\kappa_B$), derived from the Boltzmann equation within the relaxation time approximation and implemented in MUSIC \cite{Denicol:2018wdp}:
\beq \kappa_{B} = \frac{C_B}{T} n_{B} \left[ \frac{1}{3} \coth{\left(\frac{\mu_B}{T}\right)} - \frac{n_B T}{\epsilon + p} \right]. \eeq
Here, $C_B$ is a model parameter that governs the strength of baryon diffusion in the medium. In this expression, $n_B$ represents the net baryon density, $p$ is the local pressure, $T$ is the temperature, and $\mu_B$ denotes the baryon chemical potential of the fluid.

From the hydrodynamic evolution, the freeze-out hypersurface is obtained by applying a freeze-out energy density of $\epsilon_f = 0.26$ GeV/fm$^3$. The particlization is performed on this hypersurface using the Cooper-Frye prescription. Subsequently, the primordially produced hadrons undergo decays to obtain the final phase-space distribution of the produced particles. We find that the afterburner has a minimal impact on the directed flow of hadrons, which is the primary observable of interest in this study, at the considered collision energy of $\sNN = 200$ GeV. To reduce computational time and reduce statistical uncertainties, we do not employ hadronic transport to simulate interactions in the dilute hadronic phase. Instead, we use the MUSIC particlization and resonance decay routine, which provides a probability distribution for the invariant yield of each hadron species on a ($y, p_T, \phi$) grid. These distributions are then used to calculate the relevant observables.

To incorporate the effects of the electromagnetic (EM) field in our model calculations, we first determine the spatiotemporal evolution of the electric ($\vec{E}$) and magnetic ($\vec{B}$) field components generated by spectator nucleons. This is done by solving Maxwell’s equations in a conducting medium with a constant electrical conductivity $\sigma$. Subsequently, in the rest frame of each cell of the freezeout hypersurface, we compute the charge ($q$)-dependent drift velocity by solving the following force balance equation \cite{Gursoy:2014aka}:
\beq
q \vd \times \vec{B}^{\prime} + q \vec{E}^{\prime} - \mu m \vd = 0
\eeq
where $\vec{E}^{\prime}$ and $\vec{B}^{\prime}$ are the electric and magnetic fields in the fluid rest frame. The last term represents the drag force, with $\mu$ being the drag coefficient, which counteracts the Lorentz force on a charged fluid cell of mass $m$, ensuring a stationary current. This non-relativistic form of the force balance equation is justified under the assumption that the drift velocity remains much smaller than the background fluid velocity, $u^{\mu}$. In our calculations, we take $\mu m = \frac{\pi \sqrt{6 \pi}}{2}T^2$ as used in Ref. \cite{Gursoy:2018yai}. The computed drift velocity, $\vd$, has opposite signs for positively and negatively charged particles.

Next, in each fluid cell, $\vd$ is boosted by the background fluid velocity ($u^{\mu}$) to transform it back into the laboratory frame, yielding the updated drift velocity $\vec{V}$. This $\vec{V}$ represents the relativistic addition of the background velocity $u^{\mu}$ and the charge-dependent drift velocity $\vd$ induced by the electromagnetic field. We then use $\vec{V}$ in place of $u^{\mu}$ in the equilibrium distribution function within the Cooper-Frye prescription to obtain the phase space distribution of hadrons with different charges. Finally, incorporating resonance decays, we determine the final momentum space distribution of hadrons, which is used to compute the directed flow.

In this study, we do not account for the contribution of participant nucleons to the generated electromagnetic (EM) field. Previous studies have shown that the influence of participant generated EM field on $v_1$ is minimal \cite{Gursoy:2014aka}. Since this work focuses exclusively on the directed flow observable ($v_1$), we disregard the influence by participant charges. Additionally, we set $E_z = 0$, thereby neglecting any effects associated with the $z$-component of the electric field \cite{Gursoy:2014aka}.

\section{Effect of Baryon Diffusion and Electrical Conductivity}

We first analyze Au+Au collisions at $\sNN=200$ GeV to study the effects of baryon diffusion and electrical conductivity on charge-dependent directed flow splitting. The model parameters in our simulations are tuned to simultaneously describe multiple bulk observables, including the centrality and rapidity dependence of charged particle yields, the rapidity dependence of net-proton yields, and, most importantly, the directed flow of pions, protons, and anti-protons \cite{Parida:2022zse}. Our primary focus is on the centrality dependence of the $\Delta dv_1/dy$ between proton and anti-proton, as our model explicitly incorporates baryon diffusion only. Within our framework, this observable serves as a suitable probe to examine both the effects of the baryon diffusion coefficient ($C_B$) and electrical conductivity ($\sigma$).

\begin{figure}[th!]
	\includegraphics[scale=0.5]{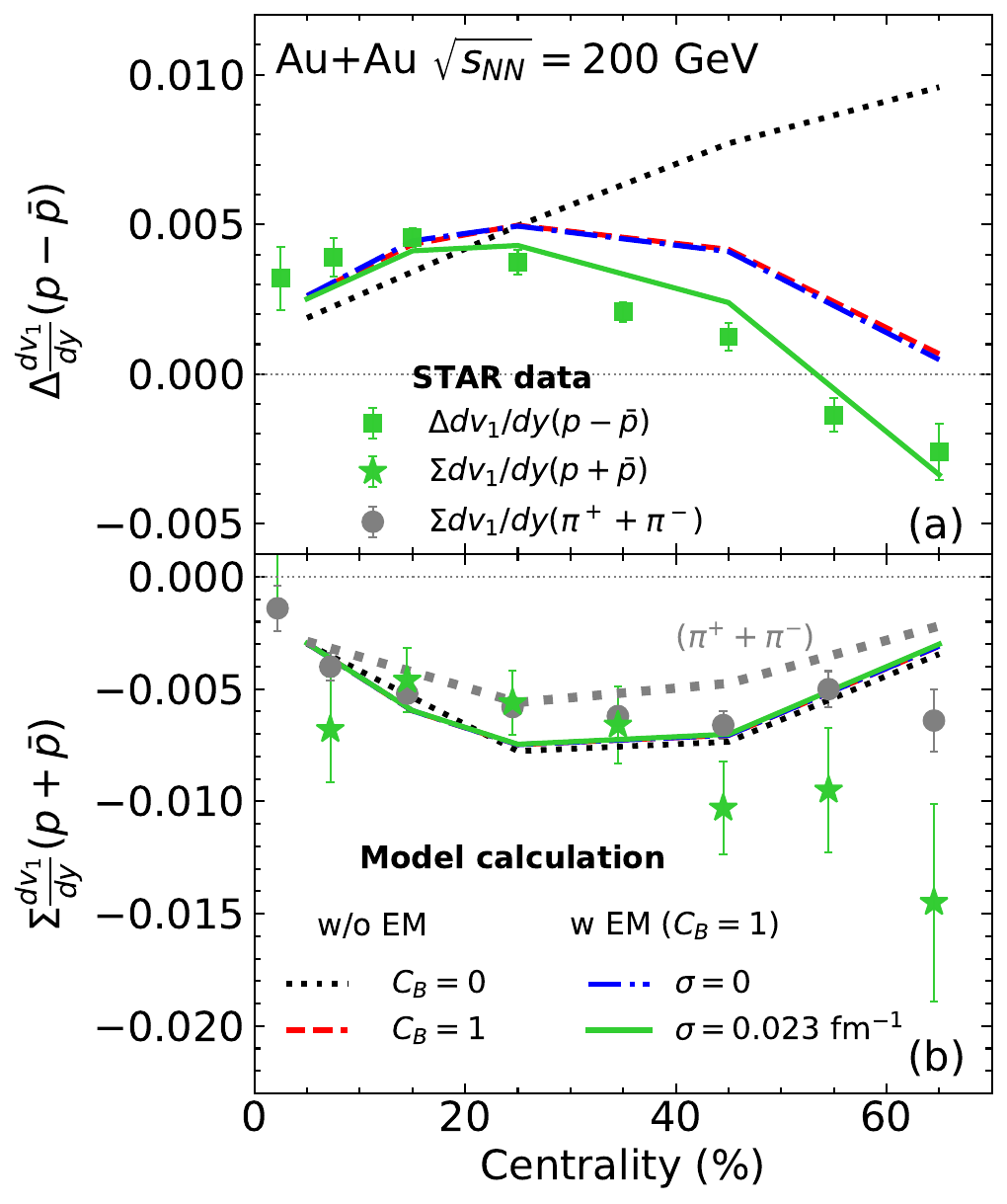} 
	\caption{ Panels (a) and (b) of this figure present centrality dependence of the difference and sum of the mid-rapidity directed flow slope between protons and anti-protons, respectively, for Au+Au collisions at $\sNN=200$ GeV. This figure illustrates the effects of baryon diffusion and electrical conductivity on $\Delta dv_1/dy (p-\bar{p})$. The dotted line represents model calculations performed without baryon diffusion ($C_B=0$) and without electromagnetic (EM) field effects. The dashed line corresponds to calculations that incorporate baryon diffusion ($C_B=1$) but still neglect the EM field. The dashed-dotted and solid lines represent cases where both baryon diffusion ($C_B=1$) and EM field effects are included, with the difference between these lines reflecting different values of electrical conductivity. The results indicate that both baryon diffusion and electrical conductivity ($\sigma$) significantly impact $\Delta dv_1/dy (p-\bar{p})$. Since the $(p+\bar{p})$ combination carries zero net-conserved charge, it remains unaffected by variations in $C_B$ and $\sigma$. To further emphasize this, panel (b) also includes the sum of $dv_1/dy$ for the $\pi^+ + \pi^-$ combination, which also has zero net-conserved charge. The centrality dependence of $\Sigma dv_1/dy (p+\bar{p})$ closely follows that of $\Sigma dv_1/dy (\pi^+ + \pi^-)$, with the only difference arising from the mass disparity between protons and pions. Experimental data from STAR are shown as symbols for comparison \cite{STAR:2023jdd,STAR:2011hyh}. The square symbols represent the data from Ref. \cite{STAR:2023jdd}, while the other two symbol types correspond to data from Ref. \cite{STAR:2011hyh}. Notably, Ref. \cite{STAR:2011hyh} provides $dv_1/dy$ for charged pions but does not explicitly report $\Sigma dv_1/dy (\pi^+ + \pi^-)$. Given that the directed flow splitting between $\pi^{+}$ and $\pi^{-}$ is expected to be minimal at $\sNN=200$ GeV, we approximate $\Sigma dv_1/dy (\pi^+ + \pi^-)$ as twice the $dv_1/dy$ of charged pions and present it for comparison in panel (b) of this figure.  } 
	\label{fig:CB_sigma_dep_cent}
        \end{figure}

\begin{figure}[th!]
	\includegraphics[scale=0.5]{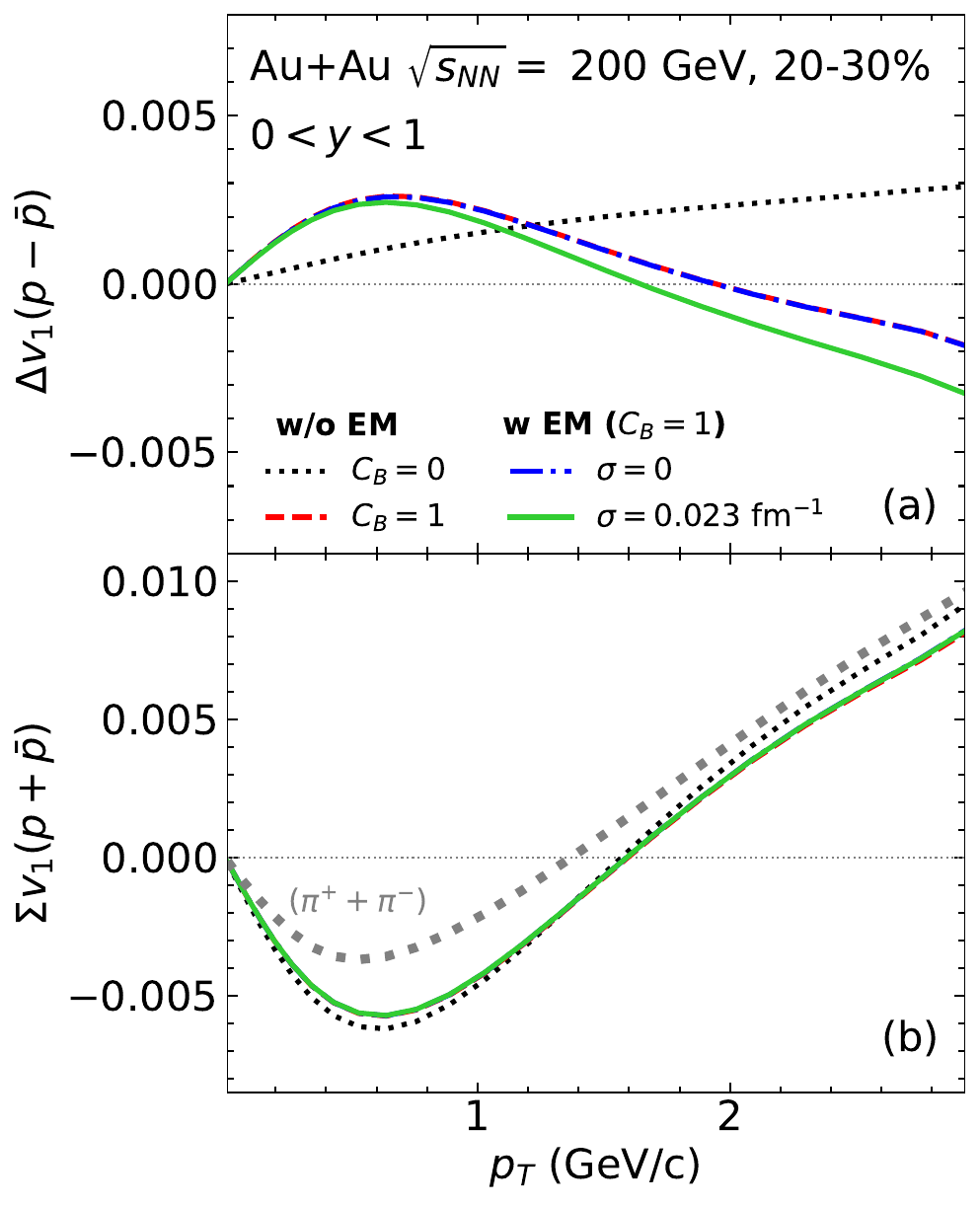} 
	\caption{ Same as Fig. \ref{fig:CB_sigma_dep_cent}, but for (a) the transverse momentum ($p_T$) differential directed flow difference between protons and anti-protons and (b) the $p_T$ differential directed flow sum. Panel (b) also includes the $p_T$ differential $v_1$ sum for $\pi^+$ and $\pi^-$ for comparison. } 
	\label{fig:CB_sigma_dep_v1_pT}
        \end{figure}

The model calculations for the centrality dependence of $\Delta dv_1/dy (p - \bar{p})$ are presented in Fig.~\ref{fig:CB_sigma_dep_cent}(a). In the absence of baryon diffusion ($C_B = 0$) and without incorporating electromagnetic (EM) field effects, $\Delta dv_1/dy (p - \bar{p})$ remains positive across all centralities. Introducing baryon diffusion with $C_B = 1$ significantly alters $\Delta dv_1/dy (p - \bar{p})$. The strong influence of baryon diffusion on the centrality dependence of $\Delta dv_1/dy (p - \bar{p})$ has been previously reported in our study at $\sNN=27$ GeV \cite{Parida:2023ldu}. It is important to emphasize that for both values of $C_B$, the model parameters, including the matter and baryon tilt parameters ($\eta_m, \omega$), are independently tuned to reproduce the aforementioned bulk observables. The parameters $\eta_m$ and $\omega$ are fixed by comparing with the directed flow data of $\pi^{\pm}$, $p$, and $\bar{p}$ at 10–40\% centrality, measured by the STAR collaboration \cite{STAR:2014clz}. For $C_B=1$, we set $\eta_m=2.0$ and $\omega=0.2$, whereas for $C_B=0$, we use $\eta_m=2.0$ and $\omega=0.31$. Furthermore, in our calculations across different centralities, we take same $\eta_m$ and $\omega$.

Next, we incorporate the effects of the EM field, with the results depicted  as dashed-dotted and solid lines in Fig. \ref{fig:CB_sigma_dep_cent}(a) corresponding to electrical conductivity values of $\sigma = 0$ and $\sigma = 0.023$ fm$^{-1}$, respectively. In both cases, the baseline calculation without the EM field corresponds to the $C_B = 1$ scenario. It is observed that for $\sigma = 0$, the EM field has a negligible effect. This is because, in a medium with zero conductivity, the initially generated magnetic field from spectator nucleons decreases rapidly, making the EM field ineffective during the later stages when flow development is significant. This minimal impact of the EM field on directed flow splitting in a zero-conductivity medium has also been reported in Ref.~\cite{Gursoy:2018yai}. However, as the conductivity increases, the magnetic field persists for a longer duration, leading to a more pronounced effect on the splitting. 

Interestingly, in the 0-10\% centrality region, the results from calculations with and without the EM field are nearly identical. This indicates that the EM field strength in central collisions is relatively weak and contributes negligibly to the splitting of $v_1$ between $p$ and $\bar{p}$. In this region, the observed splitting arises primarily due to the presence of a finite baryon density in the medium. However, in semi-central and peripheral collisions, both baryon dynamics and the EM field influence the splitting between $p$ and $\bar{p}$. Specifically, at 40-50\% centrality, incorporating baryon diffusion in our model with $C_B=1$ changes the magnitude of $\Delta dv_1/dy(p-\bar{p})$ by 46\%. Further including the EM field effects, by setting $\sigma=0.023$, results in an additional 43\% change in the magnitude of $\Delta dv_1/dy(p-\bar{p})$.

In Fig.~\ref{fig:CB_sigma_dep_cent}(b), we present the sum of the mid-rapidity slopes of the directed flow for protons and anti-protons, denoted as $\Sigma dv_1/dy (p+\bar{p})$. Since the $(p+\bar{p})$ combination has zero net-baryon number and zero net-electric charge, it is insensitive to both baryon diffusion and electrical conductivity. For comparison, we also include the $\Sigma dv_1/dy$ of the $(\pi^+ + \pi^-)$ combination, which likewise has no net conserved quantum number. The results indicate that the centrality dependence of $\Sigma dv_1/dy (p+\bar{p})$ follows a similar trend as $\Sigma dv_1/dy (\pi^+ + \pi^-)$, though a clear hierarchy emerges due to the mass difference between protons and pions. Since protons are more massive, they exhibit a larger flow magnitude \cite{Bozek:2022svy}.

When comparing with STAR data \cite{STAR:2011hyh}, we find that our model fails to accurately reproduce the centrality dependence of $\Sigma dv_1/dy (p+\bar{p})$ and $\Sigma dv_1/dy (\pi^+ + \pi^-)$. This discrepancy could stem from our simplified choice of model parameters, particularly the assumption of a fixed initial time $\tau_0$ across all centralities. In reality, the thermalization time is expected to vary with centrality due to changes in the system size. We observed that $\tau_0$ significantly influences the magnitude of $v_1$ at a given $\eta_m$, and allowing $\tau_0$ to vary with centrality could improve agreement with experimental data for both $\Sigma dv_1/dy (\pi^+ + \pi^-)$ and $\Sigma dv_1/dy (p+\bar{p})$.

In Fig.~\ref{fig:CB_sigma_dep_v1_pT}(a), we present the transverse momentum ($p_T$) dependence of the directed flow splitting between protons and anti-protons in 20–30\% centrality. Similar to our previous findings at $\sqrt{s_{NN}}=27$ GeV \cite{Parida:2023rux}, the $\Delta v_1 (p-\bar{p})$ at $\sNN=200$ GeV is also significantly influenced by baryon diffusion. When baryon diffusion is absent ($C_B=0$), the splitting remains positive across the entire $p_T$ range. However, when baryon diffusion is included ($C_B=1$), $\Delta v_1 (p-\bar{p})$ exhibits a sign change around $p_T=2$ GeV/c, transitioning from positive to negative. Comparing this result with Fig.\ref{fig:CB_sigma_dep_cent}(a), we note that at 20–30\% centrality, both $C_B=0$ and $C_B=1$ yield nearly identical magnitudes of $\Delta dv_1/dy (p-\bar{p})$. However, examining the $p_T$ dependence of $\Delta v_1 (p-\bar{p})$ in Fig.\ref{fig:CB_sigma_dep_v1_pT}(a) reveals a striking difference between the two cases.

Additionally, the electromagnetic (EM) field influences the $p_T$-dependent $\Delta v_1 (p-\bar{p})$, with its impact being more pronounced at higher $p_T$, particularly for $p_T > 2$ GeV/c. This effect of the EM field on the $p_T$-differential charge-dependent $v_1$ splitting has been previously reported in Ref.~\cite{Gursoy:2018yai}. It is important to emphasize that we focus here on the 20–30\% centrality range, where the EM field strength is relatively weaker than in more peripheral collisions, making baryon diffusion the dominant effect on $\Delta v_1 (p-\bar{p})$. Therefore, precise measurements of $p_T$-differential $\Delta v_1 (p-\bar{p})$ in central collisions, along with model-to-data comparisons in the $p_T < 2$ GeV/c region, could provide valuable constraints on the baryon diffusion coefficient $C_B$.

The transverse momentum ($p_T$) dependence of the summed directed flow, $\Sigma v_1 (p+\bar{p})$, is shown in Fig.\ref{fig:CB_sigma_dep_v1_pT}(b). Similar to the mid-rapidity slope $\Sigma dv_1/dy (p+\bar{p})$ in Fig.\ref{fig:CB_sigma_dep_cent}(b), the $p_T$-differential $\Sigma v_1 (p+\bar{p})$ remains independent of both the baryon diffusion coefficient ($C_B$) and the electrical conductivity ($\sigma$). For comparison, we also include the $p_T$ dependence of $\Sigma v_1 (\pi^+ + \pi^-)$ in the same figure. This comparison highlights that both zero net-conserved charge combinations, $(p+\bar{p})$ and $(\pi^+ + \pi^-)$, exhibit a similar $p_T$ dependence in their directed flow.

Our analysis reveals that both the baryon diffusion coefficient ($C_B$) and electrical conductivity ($\sigma$) significantly impact the centrality dependence of $\Delta dv_1/dy (p - \bar{p})$ as well as the transverse momentum dependence of $\Delta v_1 (p - \bar{p})$. This indicates that baryon diffusion constitutes a considerable background in the extraction of $\sigma$ from these observables. Therefore, a precise quantitative estimation of this background is crucial for accurately determining $\sigma$ in future studies. This can be achieved by constraining $C_B$ through model-to-data comparisons of other relevant observables. Once $C_B$ is reliably determined, it can serve as a well-defined baseline, enabling the extraction of the electromagnetic field signal.

Furthermore, in central collisions, where the EM field effects are weaker, the impact of baryon diffusion is dominant. In contrast, at peripheral collisions, both $C_B$ and $\sigma$ play a significant role. This suggests that a Bayesian analysis incorporating the centrality dependence of $\Delta dv_1/dy (p - \bar{p})$ could provide a systematic approach to constraining both $C_B$ and $\sigma$, a task we leave for future work. On the other hand, the sum of $v_1$ between protons and anti-protons remains insensitive to both baryon diffusion and electromagnetic field effects, making it a useful observable for constraining bulk parameters such as the tilt parameter and initial time.

\section{System Size Dependence}




To examine the impact of baryon dynamics and the electromagnetic field on the system size dependence of $\Delta dv_1/dy$, we conducted simulations for Ru+Ru, Au+Au, and U+U collisions at $\sNN=200$ GeV using our model.

In our calculations, we construct an event-averaged profile of participants and spectators, which serves as the basis for both the initial conditions of the hydrodynamic evolution and the spatiotemporal evolution of the EM field. We then perform a single-shot hydrodynamic evolution of this event-averaged profile for a given centrality, a method commonly employed in previous studies that has been shown to provide reasonable predictions of $v_1$ \cite{Bozek:2022svy,Du:2022yok, Parida:2022ppj,Parida:2022zse}. However, for deformed nuclei, this event-averaging approach may oversimplify the model.

Uranium, being a deformed nucleus, introduces additional complexities. Unlike spherical nuclei, where nucleon positions are the primary source of fluctuations, deformed nuclei also exhibit event-by-event fluctuations in their nuclear orientation. Consequently, the shape of the participant profile can differ significantly between the projectile and target nuclei, leading to enhanced asymmetries. Moreover, the electromagnetic field's influence on the participant zone (i.e., the energy deposition region) varies from event to event, potentially introducing additional contributions to $v_1$ splitting. While the role of nuclear deformation in $v_1$ splitting is an intriguing topic for future studies, it is beyond the scope of this work. Here, we treat U+U as a deformed nucleus but construct smooth event-averaged profiles for participant density and initial energy deposition.

In our modeling, nucleons are sampled within the nucleus according to the Woods-Saxon density distribution, parameterized as follows \cite{Shou:2014eya,Jing:2022ehp}:

\beq
\rho (r,\theta, \phi) = \frac{\rho_0}{1 + \exp{\left( \frac{r - R_0 (1+ \beta_2 Y_{20}(\theta) + \beta_4 Y_{40}(\theta) ) }{ a } \right) }}
\eeq

where $Y_{20}(\theta)$ and $Y_{40}(\theta)$ are spherical harmonics, and $\rho_0$ is a normalization constant. The parameters $R_0$, $a$, and the deformation parameters $\beta_2$ and $\beta_4$ used in our simulations for different nuclei are listed in Table \ref{tab:param_for_nucleus}. The table also includes the mass number $\mathcal{A}$ and the atomic number $\mathcal{Z}$ for each nucleus.

\begin{table}[h]
\begin{tabular}{|p{1.3cm}|p{0.8cm}|p{0.8cm}|p{1.2cm}|p{1.0cm}|p{1.0cm}|p{1.0cm}|}
\hline 
Nucleus & $\mathcal{A}$ & $\mathcal{Z}$ & $R_{0}$(fm) &  $a$  &  $\beta_2$ & $\beta_4$  \\ \hline
Ru &  96 & 44 & 5.067 & 0.5 & 0 & 0 \\
\hline
Au & 197 & 79 & 6.37 & 0.53 & 0 & 0   \\
\hline
U &  238 & 92 & 6.86 & 0.42 & 0.265 & 0.093  \\  
\hline
\end{tabular}
\caption{Parameters used in modeling the different nuclei in our simulations. The table also includes the mass number ($\mathcal{A}$) and atomic number ($\mathcal{Z}$) for each nucleus. }
\label{tab:param_for_nucleus}
\end{table}

In these simulations, we incorporate baryon diffusion by setting $C_B=1$. The model parameter values used for $C_B=1$ to describe the rapidity dependence of charged particle and net-proton yields in Au+Au collisions at $\sNN=200$ GeV are the same as those employed in our previous study \cite{Parida:2022zse}. The values of $\eta_m$ and $\omega$ have been specified in the preceding section. For other collision systems, all parameters remain unchanged except for the matter tilt parameter ($\eta_m$). Experimental data indicate that the directed flow ($v_1$) of charged particles exhibits no system size dependence \cite{STAR:2008jgm}. To ensure consistency with these observations, we independently tune the energy tilt parameter $\eta_m$ for each collision system. This adjustment ensures that the $v_1$ of charged hadrons remains consistent across all systems within the 10–40\% centrality range, aligning with experimental data \cite{STAR:2014clz}.

Since the thermalization time is expected to vary between different collision systems, the initial time for hydrodynamic evolution ($\tau_0$) may also differ. However, in our calculations, we set $\tau_0 = 0.6$ fm for all systems, necessitating system-dependent variations in $\eta_m$ to reproduce the $v_1$ data. Alternatively, a consistent $\eta_m$ value could be used with system-dependent $\tau_0$ to achieve the same effect. A more comprehensive understanding of these parameter interdependencies requires a Bayesian analysis, which would allow for a quantitative extraction of physical parameters. However, the present study focuses on a qualitative exploration of the data’s key features. Thus, we adopt a representative parameter set that provides a reasonable qualitative agreement with experimental observations.

It is important to emphasize that the baryon tilt parameter, $\omega = 0.2$, is kept same across all collision systems. This value is chosen based on Au+Au collisions to accurately describe the directed flow ($v_1$) of identified hadrons in the 10–40\% centrality range. Furthermore, for a given collision system, both $\eta_m$ and $\omega$ remain fixed across all centralities in our simulations.

\begin{figure}[th!]
	\includegraphics[scale=0.5]{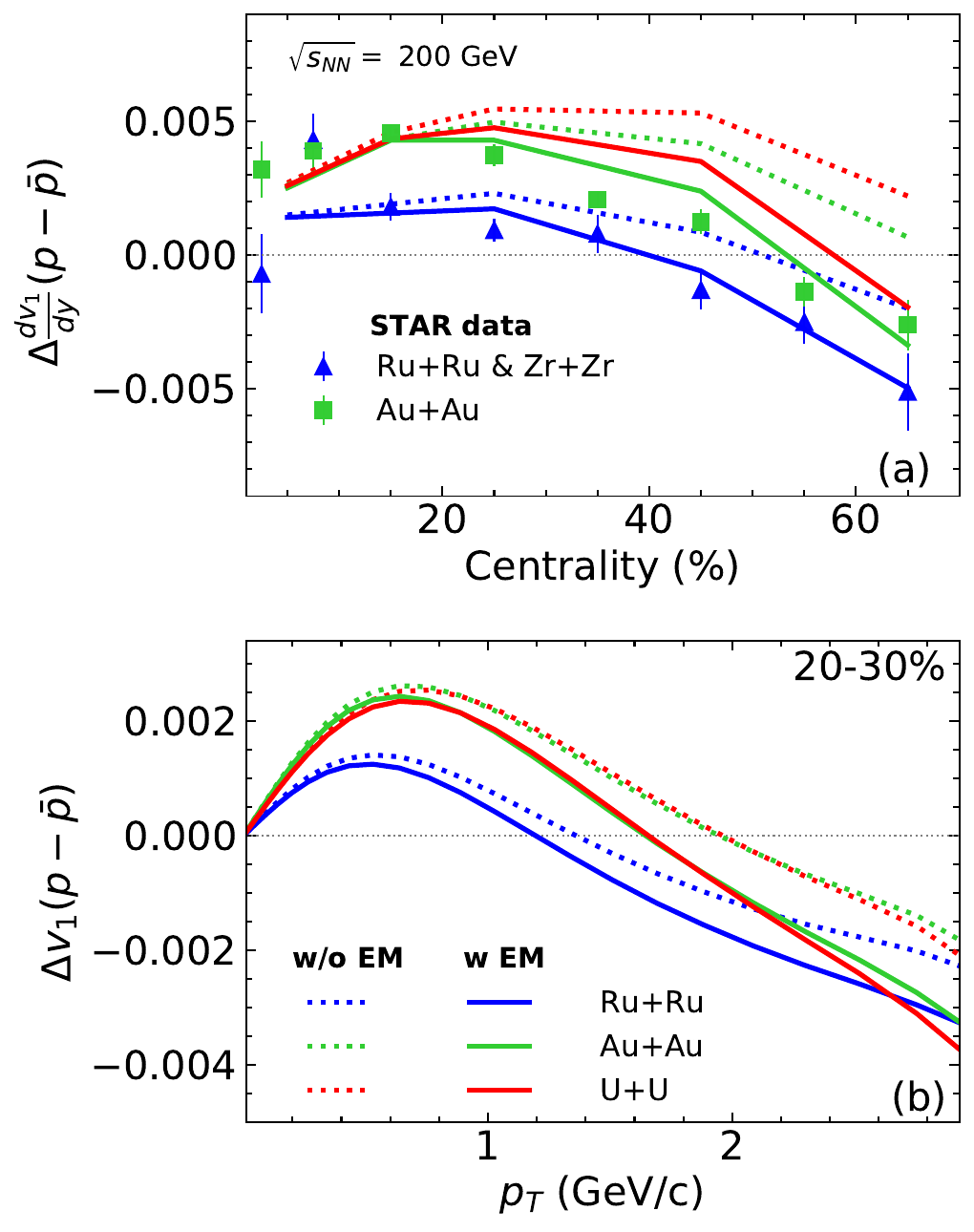} 
	\caption{ This figure illustrates the system size dependence of the $v_1$ splitting between protons and antiprotons. Panel (a): Comparison of model calculations with experimental data for the centrality dependence of the mid-rapidity slope of directed flow, $\Delta dv_1/dy (p-\bar{p})$, in Ru+Ru, Au+Au, and U+U collisions at $\sNN = 200$ GeV. Panel (b): The transverse momentum ($p_T$) dependence of the $v_1$ splitting between protons and antiprotons for the same collision systems at 20–30\% centrality. Symbols represent STAR experimental data \cite{STAR:2023jdd}, while lines denote model calculations. Different colors distinguish calculations for various collision systems. All model calculations include baryon diffusion with $C_B=1$. Dotted lines correspond to calculations without electromagnetic (EM) field effects, whereas solid lines include these effects with an electrical conductivity of $\sigma = 0.023$ fm$^{-1}$.} 
\label{fig:sys_ppbar}
\end{figure}

The simulation results for the system size dependence of the centrality differential $\Delta dv_1/dy(p-\bar{p})$ are presented in Fig. \ref{fig:sys_ppbar}(a), while Fig. \ref{fig:sys_ppbar}(b) illustrates the system size dependence of the $p_T$ differential $\Delta v_1(p-\bar{p})$ for 20–30\% centrality. Experimental data, where available \cite{STAR:2023jdd}, are shown as symbols, and our model predictions are represented by lines. The dotted lines correspond to calculations that exclude the electromagnetic (EM) field effect, whereas the solid lines incorporate the EM field with an electrical conductivity of $\sigma = 0.023$ fm$^{-1}$. Our model, which includes both baryon diffusion and EM field effects with nonzero conductivity, shows qualitative agreement with experimental observations.

In very central collisions, the EM field effect is minimal, resulting in nearly overlapping solid and dotted lines. However, our model calculations reveal a system size dependence, consistent with the expectation that larger collision systems exhibit higher baryon densities. This finding further underscores the presence of a background contribution from baryon stopping physics in $\Delta dv_1/dy (p-\bar{p})$, which complicates the extraction of the EM field signal. By comparing the dotted and solid lines across different centralities and collision systems, we observe that the EM field effect becomes more pronounced in peripheral collisions and in larger systems. Additionally, there is also a system size dependency in $p_T$ differential $\Delta v_1 (p-\bar{p})$ which could be measured experimentally.

\begin{figure}[th!]
	\includegraphics[scale=0.5]{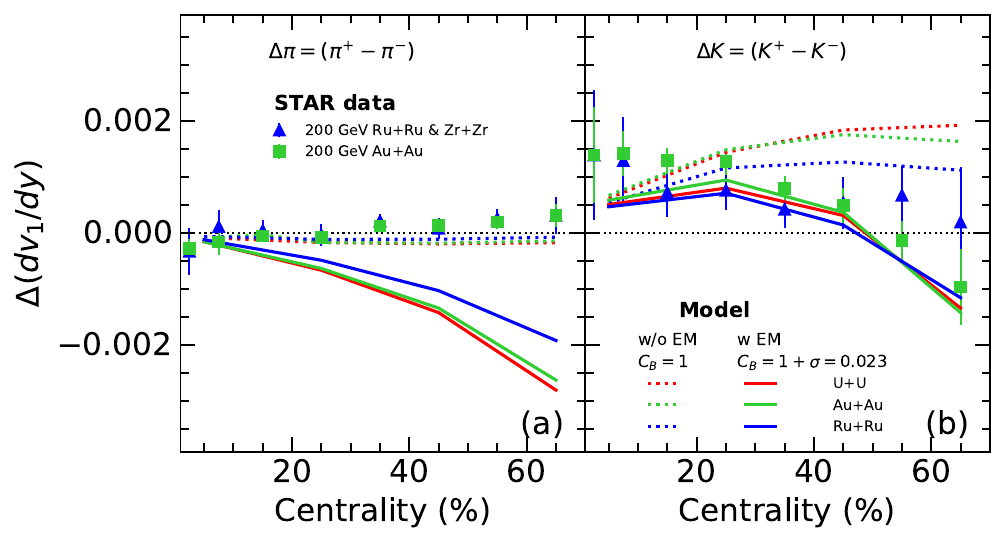} 
	\caption{ Same as panel (a) of Fig. \ref{fig:sys_ppbar}, but for (a) $\pi^{+} - \pi^{-}$ and (b) $K^+ - K^-$. } 
	\label{fig:sys_pi_k}
        \end{figure}

The system size dependence of the centrality differential $\Delta dv_1/dy$ for $\pi^+ - \pi^-$ and $K^+ - K^-$ is shown in Fig. \ref{fig:sys_pi_k}(a) and Fig. \ref{fig:sys_pi_k}(b), respectively. Our model calculations reveal a system size dependence both with and without the inclusion of EM field effects. However, due to the statistical uncertainties in the experimental data, no clear system size dependence can be confirmed. Our model successfully describes $\Delta dv_1/dy (K^+ - K^-)$ when the EM field effect is included. However, for $\Delta dv_1/dy (\pi^+ - \pi^-)$, the model fails to reproduce the measurements, as incorporating the EM field effect drives the results further away from the experimental data.

It is important to note that our model calculations consider only the initial deposition and diffusion of baryons. However, the diffusion of strangeness and electric charge, which could influence the $v_1$ splitting between $\pi^+ - \pi^-$ and $K^+ - K^-$, is not explicitly implemented. Additionally, our model does not independently deposit and evolve net strangeness and net electric charge. Instead, it employs the NEoS-BQS equation of state (EoS), which imposes local charge conservation constraints: $n_Q = 0.4 n_B$ for electric charge density and $n_S = 0$ for strangeness density \cite{Monnai:2019hkn}. Consequently, nonzero chemical potentials $\mu_S$ and $\mu_Q$ are generated in the medium, leading to a splitting in the $v_1$ of $\pi^+ - \pi^-$ and $K^+ - K^-$. Furthermore, the electromagnetic (EM) field introduces additional splitting on top of these effects. In this regard, our model is not yet a complete framework for studying the $v_1$ splitting of $\pi^+ - \pi^-$ and $K^+ - K^-$. Future model developments could enable a more detailed investigation of these observables, hence their precise interpretation remains open now. 

Additionally, studying the $v_1$ splitting between $\Lambda$ and $\bar{\Lambda}$ would be particularly interesting, as the $\Lambda$ baryon carries both strangeness and baryon number, making it sensitive to both strangeness and net-baryon diffusion effects. The correlation between the diffusion of different conserved charges could also impact the $\Delta dv_1/dy$ of $K^+ - K^-$ and $\Lambda - \bar{\Lambda}$, as these particles differ in multiple conserved quantum numbers. Future measurements and comparisons between model predictions and experimental data for these observables could provide valuable insights and open new avenues for exploration.

\section{Summary and outlook}
\label{sec:summary}

We employ a hydrodynamic framework that incorporates the effects of baryon diffusion and electromagnetic fields perturbatively, following the approach in Ref. \cite{Gursoy:2014aka,Gursoy:2018yai}. The initial energy and net-baryon distributions for the hydrodynamic evolution are taken from Ref. \cite{Parida:2022ppj, Parida:2022zse}, which successfully describes the directed flow ($v_1$) of identified hadrons, particularly baryons and anti-baryons.

Our study focuses on the interplay between baryon diffusion, controlled by the parameter $C_B$, and electrical conductivity, $\sigma$, in the centrality dependence of the mid-rapidity slope difference in directed flow between protons and anti-protons, $\Delta dv_1/dy(p-\bar{p})$, as measured by the STAR experiment \cite{STAR:2023jdd, Taseer:2024sho}. We find that both $C_B$ and $\sigma$ significantly influence the results, highlighting the necessity of accurately determining the background contribution from baryon diffusion to extract a clear electromagnetic field signal.

Additionally, we investigate the recently measured system size dependence of $\Delta dv_1/dy(p-\bar{p})$, which our model successfully reproduces. In central collisions, where the electromagnetic field strength is relatively weak, the observed system size dependence in $\Delta dv_1/dy(p-\bar{p})$ arises primarily from enhanced baryon stopping in larger systems. However, in semi-central and peripheral collisions, both baryon diffusion and electromagnetic effects contribute to $\Delta dv_1/dy(p-\bar{p})$.

Our model predictions for $\Delta dv_1/dy(\pi^+ - \pi^-)$ and $\Delta dv_1/dy(K^+ - K^-)$ are less reliable, as the model does not explicitly include the diffusion of strangeness and electric charge. These effects are expected to influence these observables in a manner similar to how baryon diffusion impacts $\Delta dv_1/dy(p-\bar{p})$. Future improvements incorporating these diffusion effects could provide a more comprehensive understanding of the observed directed flow splittings.

Although our model is not fully comprehensive, as it does not account for the complete dynamics of all conserved charges or their correlations, and treats the electromagnetic field perturbatively, it nonetheless captures several key qualitative features of the experimental data. Moreover, it provides valuable insights into the background contributions of conserved charge dynamics to electromagnetic field effects in these observables. Future advancements in hydrodynamic modeling, including the proper incorporation of all conserved charge dynamics and the implementation of magnetohydrodynamic equations \cite{Nakamura:2022wqr,Mayer:2024kkv,Mayer:2024dze}, will be essential for conducting a more precise quantitative study of these observables

\section*{Acknowledgment}
TP acknowledges the financial support from the Institute of Modern Physics, Chinese Academy of Sciences, Lanzhou, China, during his visit.

\bibliography{hydr.bib}

\end{document}